\documentclass[prd,twocolumn,letterpaper,superscriptaddress]{revtex4}
\include{epsf}
\usepackage{graphicx}
\usepackage{pslatex}

\begin{document}

\title
{Planet-sized Detectors for Ultra-high Energy Neutrinos \& Cosmic Rays
\footnote{White paper submitted for inclusion in the 
NASA Advanced Planning Office's {\it Capability Roadmap} Public Workshop, Nov. 30, 2004.}}

\author{Peter W. Gorham}
\affiliation{Dept. of Physics \& Astronomy, Univ. of Hawaii at Manoa, 2505 Correa Rd.
Honolulu, HI, 96822~~gorham@phys.hawaii.edu}

\begin{abstract}
Extragalactic astronomy with photons ends at ~0.1 PeV, but we
know there are astrophysical sources for seven more decades of energy beyond this. To
probe the highest energy sources and particles in the universe,
new messengers, such as neutrinos, and detectors with planet-sized
areas are required. This note provides a glimpse of the possibilities.
\end{abstract}

\maketitle

\begin{figure} %
%scalebox{xscale}[yscale]{\includegraphics*[x0,y0][xmax,ymax]{fig.eps}}
\scalebox{0.5}[0.4]{\includegraphics{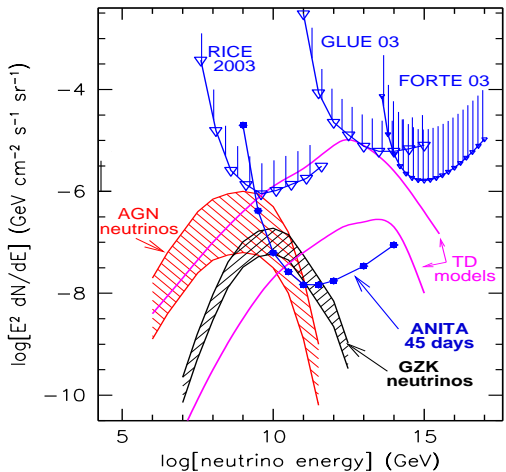}}
%\epsfxsize=3.in
%\epsfbox{Lim04.eps}
%\epsfigure{file=nulimits.eps,height=3.5in}
\caption{Model neutrino 
fluxes, including cosmogenic neutrinos, Active Galactic
Nuclei (AGN), and early universe relics 
(topological defects, TD)~\cite{GZKnu,Man96,Steck91,Yos97}. Also plotted are current
best limits, all from radio detection experiments~\cite{GLUE04,FORTE03,RICE},  
and capabilities for the NASA ANITA mission~\cite{ANITA}.}
\label{limits}
\end{figure}

%\section{Introduction}
{\bf 1. Introduction.}~~The Askaryan effect, first described in 1962~\cite{Ask62}, 
in which selective scattering and absorption processes
in a high energy particle cascade lead to a net negative charge excess and 
resulting coherent radio Cherenkov emission, has now been conclusively
observed in a Stanford Linear Accelerator (SLAC) experiment ~\cite{Sal01}. 
This result has led to renewed interest in the wider particle and
astroparticle physics communities in the technique of radio-frequency
detection of cascades from high energy neutrinos and other high energy 
particles of cosmic origin~\cite{zhsa}. In the four years since the confirmation of
this process, a series of both satellite~\cite{FORTE03} and 
ground-based experiments~\cite{RICE,GLUE04}
have now yielded the best current limits on neutrino fluxes in
hitherto unexplored energy regimes from $10^{15}$~eV to $10^{24}$~eV.
Planned experiments and missions, including the NASA-sponsored Antarctic Impulsive
Transient Antenna (ANITA) Long Duration Balloon (LDB) Mission~\cite{ANITA} will
extend the sensitivity 2-3 orders of magnitude from current levels,
and may be expected to achieve detection of the baseline cosmogenic 
ultra-high energy neutrino flux~\cite{GZKnu} at EeV ($10^{18}$~eV) energies,
an important milestone in establishing the origin of the highest
energy particles in the universe. A summary of the current state of the
field is shown in Fig.~\ref{limits}.

One of NASA's primary goals, the establishment of 
telescopes and observatories in the space or near-space environment
when such environments are fundamentally required for the exploration of our universe,
is well-served by these efforts. In fact, we cannot afford to ignore
the development of capabilities for astronomical observations above
$10^{15}$~eV (1 PeV) by a simple and compelling argument: we know there are sources
of particles of energies up to $10^{21}$~eV (1 ZeV), and our traditional messengers
for the unbiased identification and characterization of such sources--photons--cannot
propagate beyond the very local universe, a few tens of Mpc at most. This is the
result of the interaction $\gamma\gamma\rightarrow e^-e^+$, where 
first the infrared (IR) and then 3K cosmic microwave background photons provide
pair-production targets for the high energy gamma-rays. Thus
astronomy that is confined to the electromagnetic (EM) spectrum cannot directly observe
the highest energy sources in the universe, over a range of order 7 decades
of energy. To illustrate the relative magnitude of this exclusion,
one might ask the question: where would our understanding of current extragalactic
sources be if we were to remove observations over any 7 decades of the current
electromagnetic spectrum, for example, from the near IR up to MeV photons,
or from meter wavelengths to the near IR?

%\section{Ultra-high energy astronomy.}
{\bf 2. Ultra-high Energy Neutrino Astronomy.}~~
PeV to ZeV astronomy thus requires new neutral, unattenuated
messengers, with neutrinos the leading
candidate. It also requires radically new techniques for detectors and telescopes. 
At these energies the cubic-kilometer-scale of target volumes planned for
lower-energy neutrino telescopes currently under construction falls far short of
what is needed. The more relevant scale is a Teraton, or 1000 km$^3$ 
water-equivalent mass. When particles of these energies interact in such 
volumes they produce sub-nanosecond highly polarized
radio impulses of macroscopic energies, with kilowatt
peak powers at the high end of the spectrum;
such emission is detectable from even orbital distances. Thus
if the interaction media are transparent over some range of the secondary
emission spectrum, particle detection and estimation of the arrival direction 
are possible. 

The current target medium of choice is polar ice, with areas
exceeding $10^7$~km$^2$ on earth. In the 0.1-1GHz range,
cold ice is likely the most transparent solid on earth, with attenuation lengths
exceeding 1 km. From 37km balloon altitudes, $\geq 10^6$~km$^2$
of ice can be monitored; from low-earth orbit (LEO) and above, the entire Antarctic
or Greenland ice sheet is simultaneously observable. The challenge for such instruments
is thus not so much the attainment of sufficient monitored target volume,
but rather to achieve maximal sensitivity in the presence of the 240K thermal noise
levels of the ice. In this area the high level of development of antenna and
radio-frequency amplifier and receiver technology is a tremendous benefit.
In addition, the development of  unmanned aerial vehicles (UAVs) or balloons with
trajectory control that could keep station above the deepest portions 
of the Antarctic ice sheet at altitudes of 20-40 km would greatly enhance
experiments such as ANITA, which are limited primarily by dwell time above
deep ice. With station-keeping, several orders of magnitude improvement
in sensitivity are possible.

{\bf 3. Ice Planets as Detectors.}~~
Looking forward over the next several decades, we can envision the passive
use of large bodies of solar-system ice (water or hydrocarbons) as target masses
for exploration of what can be termed the high energy 
{\em electro-weak spectrum}, as particle
physicists generalize it in the current theories that unify the neutrino
weak interactions with the EM spectrum. 
Although the
surface irregularities of bodies such as Europa pose a complication to the
characterization of events, it is a tractable one, and will become more
refined as the exploration and mapping of the bodies themselves improves.
We can expect that if bodies of water ice several km deep or more are
confirmed among the
Jovian or Saturnian moons, their extremely low temperature 
(90K typical) will result in unmatched radio clarity 
compared even to the remarkable
transparency of Antarctic ice, as the microwave data from Europa
measurements already indicate~\cite{Titan03}. An all-sky neutrino mapper,
using a planetary moon such as Europa for the telescope, is 
a vision perhaps not as far-flung as it first sounds, particularly
if the thick-shell models for Europa (10-20~km ice layer) are correct. 

The information content of the secondary RF emission from a cascade
is distributed in the polarization angle vs. location along the radio
Cherenkov cone; in the intrinsic radio spectrum observed, and its
spatial gradient; and in the time structure of the impulse.
To improve the detection signal-to-noise ratio and the
measurement precision, a close (several km) formation of several 
satellites, able to perform polarimetry and pulse-phase interferometry
over more than a single location of the Cherenkov cone of the emission would 
have significant advantages for any of the methods envisioned here.

{\bf 3. Cosmic Ray Detection from Orbit.}
Several proposed missions~\cite{OWL,EUSO} have already advanced 
the concept of extending ground-based ultra-high energy cosmic ray 
optical fluorescence detection to an earth-orbiting platform,
to observe cosmic ray air showers over very wide areas of
the earth's atmosphere ($\sim 10$~m water equivalent).
Other planetary surface materials, such as the lunar regolith, 
while not as radio-transparent
as ice, can still be observed to depths of several tens of m, and thus
even a lunar orbiter may have access to many Teratons of target material.
In addition, the lack of an atmosphere on objects such as the Moon mean
that other particle species that cascade close to the surface, such as
the ultra-high energy cosmic rays, are also in principle observable, since
they also produce coherent radio emission via the Askaryan process.
For example, an orbiter (or again a compact formation-flying satellite
group) at 1 lunar radius above the Moon's surface
would synoptically observe $10^7$~km$^2$ of regolith, over which
there is of order $10^4$ ultra-high energy cosmic ray events per year
above $10^{21}$~eV. Such events, each dumping more than 160 Joules of
electromagnetic energy into the
top few m of regolith, will produce strong coherent RF
impulses which should be straightforward to detect at the maximum
3000 km distance from the horizon to such an orbiter or formation. This principle
has already been applied to searches for ZeV neutrino interactions in the
lunar regolith by the GLUE experiment and others~\cite{GLUE04,Zhe88}.

{\bf 4. Conclusions.}~~
The latter half of the twentieth century 
has seen NASA develop the means to 
explore the solar system and beyond using direct spacecraft travel,
and orbiting telescopes created to exploit the pristine environment of space. 
If we are to further our exploration of the frontier of 
ultra-high energies to its currently known limits, we
will need to move beyond the pure electromagnetic spectrum, and
synthesize new kinds of telescopes, some the size of entire planets.
This methodology draws upon the same kind of vision that led to
the 20th century confirmation of the theory of general relativity, using the
gravitational influences of the sun and planets: 
to see the planetary system itself
as a rich scientific resource enabling us to look far beyond its
limits, leading to entirely
new kinds of astronomy, and new windows on the extreme universe.

%\end{references}
 

\begin{thebibliography}{99}

\bibitem{Ask62} G. A. Askaryan,1962, JETP 14, 441.
\bibitem{Sal01} D.  Saltzberg, P. Gorham, D. Walz, {\it et al.} Phys.
Rev. Lett., {\bf 86}, 2802 (2001); hep-ex/0011001.
\bibitem{zhsa} E. Zas, F. Halzen, \& T. Stanev, 1992, Phys Rev D 45, 362;
J. Alvarez--Mu\~niz, \& E. Zas, 1997, Phys. Lett. B, 
411, 218.
\bibitem{FORTE03} N. Lehtinen, P. Gorham, A. Jacobson, \& R. Roussel-Du\'pre,
Phys.Rev. D69 (2004) 013008 ; astro-ph/0309656.
\bibitem{RICE} I. Kravchecnko et al., Astropart.Phys. 20 (2003) 195-213.
\bibitem{GLUE04} P.W. Gorham et al., Phys. Rev. Lett. 93 (2004) 041101; astro-ph/0310232.
\bibitem{ANITA} A. Silvestri et al. (2004) astro-ph/0411007;
\url{http://www.phys.hawaii.edu/~anita}
\bibitem{GZKnu}R. Engel et al. PRD 64, 093010, (2001).
\bibitem{Yos97} S. Yoshida et al. 1997, ApJ 479, 547; Bhattacharjee, P. et al., 1992 PRL 69, 567.
\bibitem{Titan03} R. D. Lorenz, et al., Planet. Space Sci. 51 (2003), 353.
\bibitem{Zhe88} R. D. Dagkesamanskii, \& I. M. Zheleznyk, 1989, JETP 50, 233;
T. H. Hankins et al., 1996, MNRAS 283, 1027.
\bibitem{Man96} Mannheim, K., 1996, Astropart. Phys 3, 295.
\bibitem{Steck91}F.W. Stecker et al.,PRL 66,2697 (1991).
\bibitem{OWL}OWL: Orbiting Wide-area Light collectors, F. W. Stecker et al.(2004),
Nucl. Phys. B, in press; astro-ph/0408162.
\bibitem{EUSO}EUSO: ``Extreme Universe Space Observatory'' O. Catalano,
Il Nuovo Cimento, Vol.24-C, N.3, pp.445-470, 2001.




\end{thebibliography}
\end{document}